\documentclass[twocolumn,showkeys,preprintnumbers,amsmath,amssymb,prd]{revtex4}
\usepackage[dvips]{graphicx}
\usepackage{dcolumn}
\usepackage{bm}
\begin{document}


\title{The Mario Schenberg Gravitational Wave Detector:\\A mathematical model for its quadrupolar oscilations}

\author{C\'{e}sar A. Costa}
\email{cesar@das.inpe.br}
\author{Odylio D. Aguiar}
\email{odylio@das.inpe.br}
\affiliation {Divis\~ao de
Astrof\'{i}sica, Instituto Nacional de Pesquisas Espaciais \\C.P.
515, S\~ao Jos\'{e} dos Campos, SP, 122201-907, Brazil}

\author{Nadja S. Magalh\~aes}
\email {nadjam@ig.com.br} \affiliation { Instituto
Tecnol\'{o}gico de Aeron\'{a}utica, Departamento de F\'{i}sica \\
Pra\c{c}a Mal. Eduardo Gomes
50, S\~ao Jos\'{e} dos Campos, SP, 12228-900, Brazil \\
}

\date{\today}

\begin{abstract}
In this work we present a mathematical model for the mechanical
response of the Brazilian Mario SCHENBERG gravitational wave (GW)
detector to such waves. We found the physical parameters that are
involved in this response assuming a linear elastic theory.
Adopting this approach we determined the system's resonance
frequencies for the case when six $i$-mode mechanical resonators
are coupled to the antenna surface according to the arrangement
suggested by Johnson and Merkowitz: the truncated icosahedron
configuration. This configuration presents special symmetries that
allow for the derivation of an analytic expression for the mode
channels, which can be experimentally monitored and which are
directly related to the tensorial components of the GW. Using this
model we simulated how the system behaves under a gravitational
sinewave quadrupolar force and found the relative amplitudes that
result from this excitation. The mechanical resonators made the
signal $\approx 5340$ times stronger. We found $i+1$ degenerate
triplets plus $i$ non-degenerate system mode resonances within a
band around ${3.17-3.24\rm{kHz}}$ that are sensitive to signals
higher than ${\tilde h\sim 10^{-22}\rm{Hz}^{-1/2}}$ when we
considerate the effects of thermal noise only.
\end{abstract}

\maketitle

\section{\label{sec:intro}Introduction}

The existence of gravitational waves (GWs) was predicted by
Einstein applying his General Relativity Theory to the vacuum in
the weak field approximation \cite{einstein1916}. GWs are local
space-time curvature perturbations caused by accelerated masses.
These perturbations travel through spacetime with the speed of
light and can excite quadrupolar normal-modes of elastic bodies.
\par Weber was the first one to propose feasible gravitational wave
detectors, 43 years ago \cite{weber1960}. However, so far nobody
was able to convince the scientific community of any detection.
The observable evidence of the existence of these waves comes from
an indirect observation in the electromagnetic spectrum of a
pulsar radio system \cite{taylor1994}.
\par In 1971 Forward suggested the use of a sphere as the antenna
element of a resonant-mass detector \cite{forward1971}. He
idealized a sphere with nine sets of electromechanical strain
transducers placed along great circle routes. The tensor
gravitational radiation components would be determined by five
independent quantities from the nine transducer outputs. Ashby and
Dreitlein studied in detail the reception of GWs by an elastic
self-gravitating spherical antenna \cite{ashby1975} and Wagoner
and Paik found the lowest eigenvalues for the monopole and
quadrupole modes of a uniform elastic sphere \cite{wagoner1976}.
In the 1990's, Johnson and Merkowitz studied the
antenna-transducer coupling problem and found an optimum
configuration, which minimizes the number of transducers while
keeping them in a symmetric distribution on the antenna: the
truncated icosahedron (TI) configuration \cite{johnson1993}.
Magalhaes and collaborators showed that distributions with more
resonators also can have interesting symmetric properties
\cite{magalhaes1997}.
\par In this work we extend Johnson and Merkowitz's model to two-mode
mechanical transducers. This kind of transducer will be used on
the \textbf{Mario SCHENBERG} spherical detector, which is being
constructed in the Physics Institute of the Sao Paulo University
with financial support from the State of Sao Paulo Research
Foundation (FAPESP).
\begin{figure}\label{fig:schenberg}
  \includegraphics[width=8.5cm]{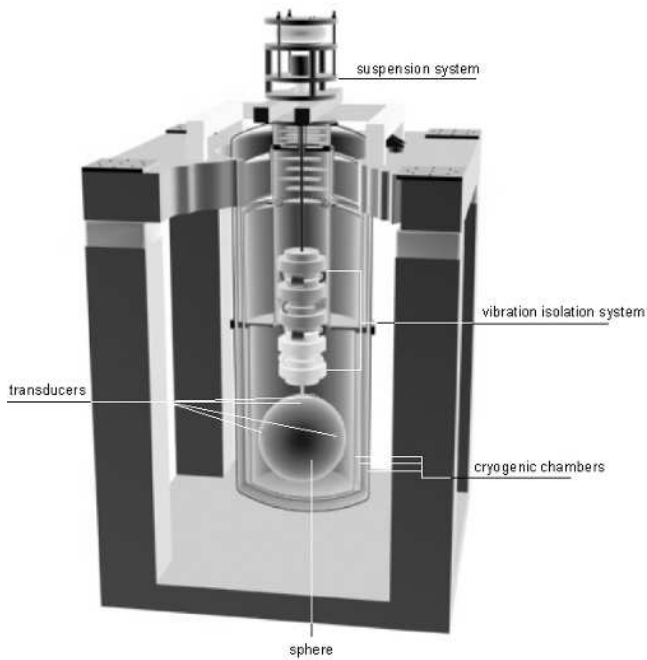}
  \caption{Schematic diagram of the Brazilian spherical GW detector Schenberg.}
\end{figure}
\par SCHENBERG has a $65\rm{cm}$-diameter and $\sim 1150\rm{kg}$
copper alloy $[\rm{Cu}(94\%)\rm{Al}(6\%)]$ spherical antenna that
will operate at temperatures below $1\rm K$. These characteristics
are shared with the detectors MiniGrail (The Netherlands) and
SFERA (Italy). Figure \ref{fig:schenberg} shows a schematic
diagram of the Schenberg detector.
\par In this work we consider an isolated sphere coupled only to
mechanical resonators and predict the mechanical behavior of the
system. We solve the displacement equations of the whole system
and find an analytic expression that relates the mode channels
measurements to the GWs tensorial components.

\section{\label{sec:sphere} The sphere}
We will assume that the antenna has a perfect spherical symmetry
and ignore the effects caused by the hole machined in the body to
attach the suspension rod. We will show in subsection
\ref{ssec:freq} by numerical simulations that this is a good
approximation for the real case.
\par Consider a spherical isotropic elastic body with radius $R$ and density
$\rho$ in a field of forces. The strain suffered by a mass element
$dM$ is described by the displacement vector ${u_i=x_i'-x_i}$
(where $x_i$ represents the equilibrium position and $x_i'$
denotes the position of $dM$ after deformation). The strain tensor
$u_{ij}$ is defined by \cite{bianchi1996}
\begin{equation}
u_{ij}=\frac{1}{2}\left(u_{i,j}+u_{j,i}\right), ~~\rm{with} ~~i,
j=1,2,3.
\end{equation}
There is a stress tensor $\sigma_{ij}$ associated to strain tensor
by constitutive equations (generalizations of Hooke's law), given
by
\begin{equation}\label{eq:sigma}
\sigma_{ij}=\delta_{ij}\lambda u_{ll}+2\mu u_{ij},
\end{equation}
where $\lambda$ and $\mu$ are Lam\`e coefficients, which describe
elastic proprieties of a body (they are functions of the Poisson's
ratio and the Young modulus). By the application of the principle
of conservation of linear momentum we find equilibrium equations
given by
\begin{equation}
\sigma_{ij,j}+\rho u_{i,00}=f_i,
\end{equation}
where $f_i$ corresponds to the $i$-th component of an external
force density field and $\rho {u_i}_{,00}$ to an internal one.
These equations are called Navier's equations and can be rewritten
by application of equation \ref{eq:sigma} as
\begin{equation}\label{eq:navier}
(\lambda +\mu)u_{k,ki}+\mu u_{i,jj}+\rho {u_i}_{,00}=f_i,
\end{equation}
The Navier's equations must be satisfied by a set of functions
$u_i=u_i(x_1,x_2,x_3)$ which represent the displacements inside a
region limited by a radius $R$. Therefore Equation \ref{eq:navier}
requires boundary conditions which will indicate that the surface
is free to oscillate, defined as
\begin{equation}\label{eq:cc}
n_j\sigma_{ij}=0, ~~\mathrm{for}~~ r=R,
\end{equation}
where $n_i\equiv x_i/r$ is the normal unit.

We assume that it is possible to separate the spacial and temporal
dependencies of the displacement vector at a position $\mathbf{x}$
in a time $t$, so we have
\begin{equation}
u_i(\mathbf{x},t)=\sum_m A_m(t)\mathbf{\Psi}_m(\mathbf{x}),
\end{equation}
where $A_m(t)$ is the vibrational amplitude of the $m$-th normal
mode and $\mathbf{\Psi}_m(\mathbf{x})$ is the eigenfunction of
this mode. In fact, if the sphere is made of ordinary matter
($v_{sound}\ll c$) its radius is much smaller than the wavelength
and its mechanical quality factor is high ($Q_m=\omega_m\tau_m\gg
1$, where $\omega_m$ and $\tau_m$ are the angular frequency and
the energy time decay of $m$ mode, respectively) such separation
can be done.
\par The eigenfunctions are normalized according to
\begin{equation}\label{eq:ccPsi}
\int_V \mathbf{\Psi}_n(\mathbf{x})\cdot
\mathbf{\Psi}_m(\mathbf{x})d^3x=N_m\delta_{mn},
\end{equation}
where $N_m$ is a arbitrary normalization factor for mode $m$. If
sphere is homogeneous (with constant $\rho$), we obtain
\begin{equation}
N_m\equiv \frac{4\pi}{3}R^3, ~~\forall~~ m.
\end{equation}
\par The equation of motion for the sphere as a forced harmonic
oscillator can be obtained from Navier's equation and it is given
by
\begin{eqnarray}\label{eq:mov1}
\ddot A_m(t)+\tau_m^{-1}\dot A_m(t)+\omega^2_0
A_m(t)=\frac{1}{\rho
N_m}\times\nonumber\\
\int_{V_0}\mathbf{\Psi}_m(\mathbf{x})\cdot\sum
\mathbf{f}(\mathbf{x},t)d^3x.
\end{eqnarray}
The eigenfunctions $\mathbf{\Psi}_m(\mathbf{x})$ lead information
about the sphere's elastic proprieties ($\lambda$ and $\mu$), as
well vibrational features (i.g. $\omega_0$).
\par
\begin{figure}
  \includegraphics[width=8.5cm]{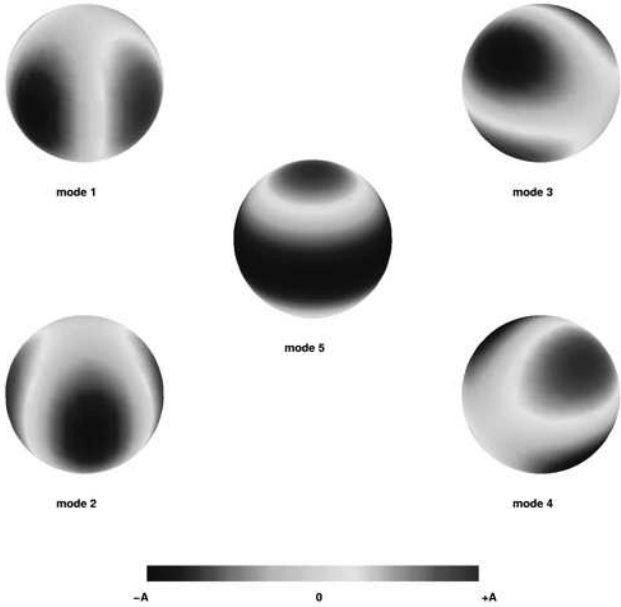}
  \caption{Quadrupolar normal modes of sphere.
  The darkest areas represent places of highest radial motion.}
  \label{fig:modes}
\end{figure}

 Specific solutions for Equation \ref{eq:mov1} can be obtained
using spheroidal modes described as linear combinations of the
quadrupolar spherical harmonics $Y_m(\theta,\phi)$ to determine
$\mathbf{\Psi}_m(\mathbf{x})$, which assumes the form
\cite{merkowitz1997}
\begin{equation}\label{eq:Psi}
\mathbf{\Psi}_m(r,\theta,\phi)=[\alpha(r)\mathbf{\hat{r}}+\beta(r)R\nabla]Y_m(\theta,\phi),
\end{equation}
where $\alpha(r)$ and $\beta(r)$ are respectively the radial and
tangential motion parameters. Figure \ref{fig:modes} shows the
distribution of the radial part of $\mathbf{\Psi}_m$ on the sphere
surface, the darkest regions representing the highest radial
amplitudes. Ashby and Dreitlein described $\alpha(r)$ and
$\beta(r)$ function as \cite{ashby1975}
\begin{eqnarray}
  \alpha(r)&=& p_1R\frac{\partial}{\partial r} j_2(qr)+6p_2\frac{R}{r}j_2(kr)~~\mathrm{and}\\
  \beta(r)&=& p_1j_2(qr)+6p_2\frac{\partial}{\partial
  r}[rj_2(kr)],
\end{eqnarray}
where $j_2(x)$ is the spherical second order Bessel's function
defined as \cite{arfken2000}
\begin{equation}
j_2(x)=\left(\frac{3}{x^3}-\frac{1}{x}\right)\sin x - \frac{3\cos
x}{x^2},
\end{equation}
and $q^2=\rho \omega^2_0/(\lambda+2\mu)$ and $k^2=\rho
\omega^2_0/\mu$ are respectively the squared transverse and
longitudinal wave vectors. The value of the eigenfrequency
$\omega_0$ is determined from Equation \ref{eq:cc} which can be
rewritten as
\begin{eqnarray}
 p_1\frac{d}{dr}\left[\frac{j_2(qr)}{r}\right]+p_2\left[\frac{10-(kr)^2}{2r^2}-\frac{1}{r}\frac{d}{dr}\right]j_2(kr)=0\nonumber,\\
 p_1\left[\frac{12-(kr)^2}{2r^2}-\frac{2}{r}\frac{d}{dr}\right]j_2(qr)+6
 p_2\frac{d}{dr}\left[\frac{j_2(kr)}{r}\right]=0\nonumber,
\end{eqnarray}
where we consider only quadrupolar modes. The values of the
normalization parameters $p_1$ and $p_2$ are found from the
normalization condition given by Equation \ref{eq:ccPsi}.
\par Since we know the values of these parameters an expression for
effective gravitational force that acts on the sphere can be found
as
\begin{equation}\label{eq:egf}
\mathcal{F}_m(t)=\frac{1}{2}\ddot{h}_m(t)m_S \chi R,
\end{equation}
where
\begin{equation}
\chi=\sqrt{\frac{3}{15\pi}}\left[p_1j_2(qR)+3p_2j_2(kR)\right]
\end{equation}
is a factor that determine the effective mass of the sphere, given
by \cite{harry1996}
\begin{equation}
m_{eff} = \frac{5}{12}\chi\left(\frac{4\pi}{3}R^3\right)\rho
~~\Rightarrow~~ m_{eff} \sim \frac{m_S}{4},
\end{equation}
and $h_m=h_m(h_{+}(t),h_{\times}(t),\theta,\varphi)$ denotes the
spherical amplitude of gravitational wave that depends on the
polarization amplitudes and the orientation $[\theta,\varphi]$
relative to the lab coordinate system centered in the sphere's
center of mass (SCM) \cite{merkowitz1998}.
\par We solved the mathematical model for the sphere and found the specific physical
parameters of the SCHENBERG detector, as is shown in Table
\ref{tab:values}.
\begin{table}
\caption{\label{tab:values}Parameters for the SCHENBERG detector:
introduced(1) and obtained from the present model(2).}
\begin{ruledtabular}
\begin{tabular}{lcc}
Description & Symbol & Value\\\hline
(1) sphere radius at 4K & $R$ & $0.3239\mathrm{m}$ \\
(1) sphere mass & $m_S$ & $1147.85\mathrm{kg}$ \\
(1) sphere effective mass & $m_{eff}$ & $287.63\mathrm{kg}$ \\
(1) mass of resonator $R_1$ & $m_{R_1}$ & $10^{-5}\mathrm{kg}$ \\
(1) mass of resonator $R_2$ & $m_{R_2}$ & $0.054\mathrm{kg}$ \\
(1) density at 4K & $\rho$ & $8065.7\mathrm{kg/m^3}$ \\
(1) Young modulus at 4K & $E$ & $1.33\times10^{11}\mathrm{Pa}$ \\
(1) Poisson's ratio & $\nu$ & $0.364$ \\
(1) linear contraction coefficient\footnotemark[1] & $\Delta R/R$ & $334.52\times 10^{-5}$ \\
(1) bulk Lam\`e's coefficient at 4K& $\mu$ & $4.8753\times10^{10}\mathrm{Pa}$\\
(1) shear Lam\`e's coefficient at 4K& $\lambda$ & $1.3049\times10^{11}\mathrm{Pa}$ \\
(1) sphere Q\footnotemark[2] & $Q_S$ & $2.0\times 10^6$\\
(1) first resonator Q & $Q_{R_1}$ & $10^6$\\
(1) second resonator Q & $Q_{R_2}$ & $10^6$\\
(2) degenerate mode frequencies & $f_0$ & $3206.3\mathrm{Hz}$ \\
(2) normalization parameter 1 & $p_1$ & $-5.5654$ \\
(2) normalization parameter 2 & $p_2$ & $2.2758$ \\
(2) radial parameter at $r=R$& $\alpha(R)$ & $2.8623$ \\
(2) tangential parameter at $r=R$& $\beta(R)$ & $0.6598$ \\
(2) Chi factor & $\chi$ & $0.6013$ \\
\end{tabular}
\end{ruledtabular}
\footnotetext[1]{Tenderized mean
($94\%\mathrm{Cu}+6\%\mathrm{Al}$) from \cite{scott1963}.}
\footnotetext[2]{Mechanical quality factor}
\end{table}
\begin{figure}
  \includegraphics[width=8.5cm]{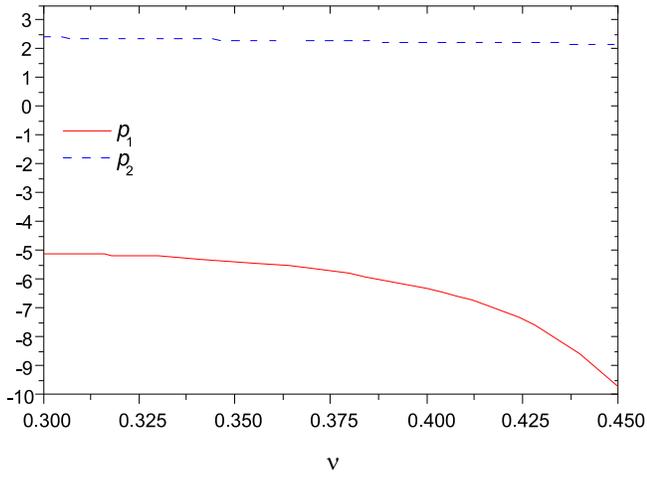}\\
  \caption{\label{fig:dependence1}The normalization parameters dependencies on the Poisson's ratio $\nu$.}
\end{figure}
\begin{figure}
  \includegraphics[width=8.5cm]{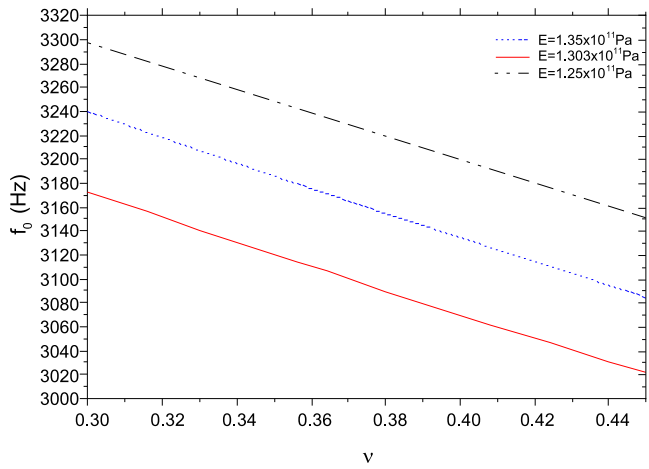}\\
  \caption{\label{fig:dependence3}The resonance frequency $f_0$ dependence on $\nu$ and on the Young modulus $E$.}
\end{figure}
\begin{figure}
  \includegraphics[width=8.5cm]{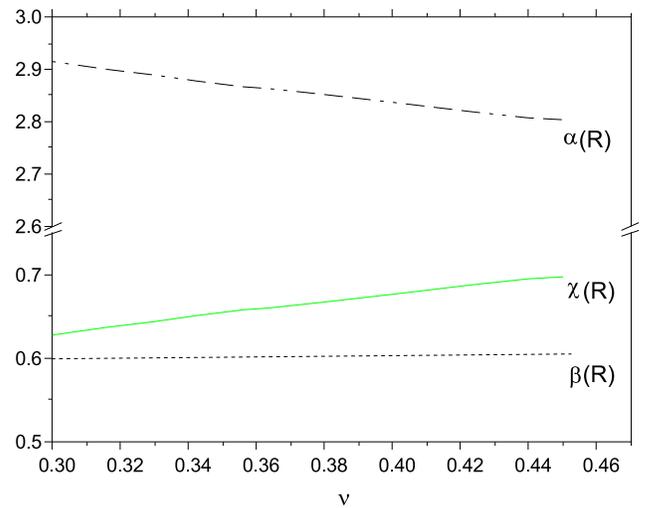}\\
  \caption{\label{fig:dependence4}The radial $\alpha(R)$ and tangential $\beta(R)$ displacement parameters
  and $\chi(R)$ factor dependencies on $\nu$ at the sphere surface ($r=R$).}
\end{figure}
The physical dependencies between parameters can be seen in
Figures \ref{fig:dependence1}, \ref{fig:dependence3} and
\ref{fig:dependence4}. It is easy to notice that $\alpha(R)$ and
$\chi(R)$ have small dependence on the Poisson's ratio. Our tests
shown that they are practically Young modulus independent. These
features imply that little discrepancies on values used in our
model will not affect hardly the results obtained in this work.

\par Evidently, the gravitational force presented in Equation
\ref{eq:egf} is not the only force that acts on the sphere. The
resultant force has other components $F_S^N$ that come from noise
sources (e.g. Langevin's forces). These forces must be taken into
consideration because they will contribute to the sum of the
forces in Equation \ref{eq:mov1}. Thus the sphere will be under
the action of the forces ${F^S}_m=\mathcal{F}_m+{F_S^N}_m$. Many
ways to minimize the noise forces contributions have been studied
and a lot of what have been learned in the last decades with bar
instruments has been reused and reorganized to be applied to
spherical detectors (e.g. vibration isolation systems,  cryogenic
devices, electromagnetic isolation, etc).
\par On the other hand, an isolated sphere is not a practical detector since
the deformation it suffers due to the GW influence is too small to
be detected. Something is needed to work as a impedance
transformer. Thus we coupled secondary resonators to the antenna
in order to amplify the signal.

\section{\label{sec:sphereplusresonators} The sphere coupled to resonators}

\begin{figure}
  \includegraphics[width=8.5cm]{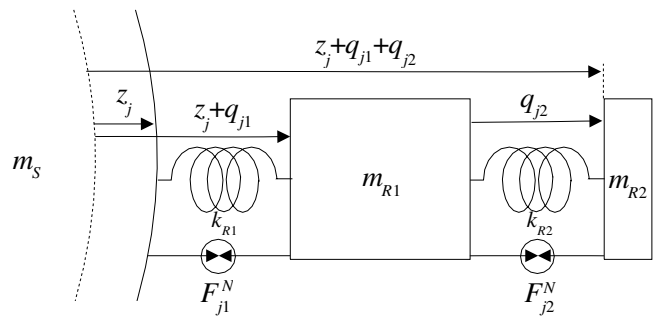}\\
  \caption{\label{fig:resonator}Unidimensional mechanical resonator with
  $2$ modes attached to the sphere's surface.}
\end{figure}

Suppose there are $j$ resonators with $2$ modes coupled to the
sphere's surface, as it is shown in Figure \ref{fig:resonator}.
The resonator is assumed to displace only in the radial direction.
When the sphere's surface suffers a radial displacement described
by the vector $z_j$ under each resonator $j$, with masses
$m_{R_1}$ and $m_{R_2}$ that will represent them henceforth, we
have
\begin{equation}
z_j(t)=\hat{\mathbf{r}}_j \cdot \sum_m A_m(t)\mathbf{\Psi}_m(t),
\end{equation}
relative to the SCM, with $\hat{\mathbf{r}}_j$ being the unit
radial vector at the $j$-th resonator's position. Thus momentum is
transferred to the $j$ resonator, making $m_{R_1}$ displace
$q_{j_1}+z_j$ from the SCM. Thus $m_{R_2}$ moves a distance
$q_{j_2}+q_{j_1}+z_j$ from that point.
\par The radial displacements of the sphere's surface under each
resonator correspondent to each  mode $m$ can be collected in $m$
vectors that describe the pattern of the radial displacement for
each mode. Those ``pattern vectors" can be grouped in a ``pattern
matrix" $B_{mj}$, defined by \cite{merkowitz1997}
\begin{equation}\label{eq:Bmj}
B_{mj}=\frac{1}{\alpha}\hat{\mathbf{r}}_j \cdot
\mathbf{\Psi}_m(x_j),
\end{equation}
where $\alpha=\alpha(R)$ represents the radial displacement
parameter value at the sphere's surface. Using Equation
\ref{eq:Psi} we can rewrite \ref{eq:Bmj} as
\begin{equation}
B_{mj}\equiv Y_m(\theta_j,\phi_j).
\end{equation}
In other words, $B_{m_j}$ represents a spherical harmonic $Y_m$
behavior at $x_j$ position $x_j$.
\par The equations of motion of the system can be written in matrix form as
\begin{eqnarray}\label{eq:moveq}
\left(%
\begin{array}{ccc}
  m_S\underline{\underline{I}} & \underline{\underline{0}} & \underline{\underline{0}} \\
  m_{R_1}\alpha\underline{\underline{B}}^T & m_{R_1}\underline{\underline{I}} & \underline{\underline{0}} \\
  m_{R_2}\alpha\underline{\underline{B}}^T & m_{R_2}\underline{\underline{I}} & m_{R_2}\underline{\underline{I}} \\
\end{array}%
\right)\left(%
\begin{array}{c}
  \underline{{\ddot{A}}} (t) \\
  \underline{{\ddot{q_1}}} (t) \\
  \underline{{\ddot{q_2}}} (t) \\
\end{array}%
\right)+\nonumber\\
\omega_0^2\left(%
\begin{array}{ccc}
  m_S\underline{\underline{I}} & -m_{R_1}\alpha\underline{\underline{B}} & \underline{\underline{0}} \\
  \underline{\underline{0}} & m_{R_1}\underline{\underline{I}} & -m_{R_2}\underline{\underline{I}} \\
  \underline{\underline{0}} & \underline{\underline{0}} & m_{R_2}\underline{\underline{I}} \\
\end{array}%
\right)
\left(%
\begin{array}{c}
  \underline{A} (t) \\
  \underline{q_1} (t) \\
  \underline{q_2} (t) \\

\end{array}%
\right)=\nonumber\\
\left(%
\begin{array}{ccc}
  \underline{\underline{I}} & -\alpha\underline{\underline{B}} & \underline{\underline{0}} \\
  \underline{\underline{0}} & \underline{\underline{I}} & -\underline{\underline{I}} \\
  \underline{\underline{0}} & \underline{\underline{0}} & \underline{\underline{I}} \\
\end{array}%
\right)\left(%
\begin{array}{c}
  \underline{F_S} (t) \\
  \underline{F_1}^N (t) \\
  \underline{F_2}^N (t) \\
\end{array}%
\right),
\end{eqnarray}
where elements double underscored are matrices and single
underscored ones are column vectors. We have
$\underline{\underline{B}}^T=[B_{jm}]$ as transpose of
$\underline{\underline{B}}=[B_{mj}]$, $\underline{\underline{I}}$
is the identity matrix and $\underline{\underline{0}}$ is the null
matrix. The dimensions of these matrices depends only of the
values of $j$ and $m$.
\par Equation \ref{eq:moveq} was obtained assuming that
the components of the system have high  mechanical quality factor
($Q>10^6$). Consequently, the damping coefficients are negligible
and the time decay of the oscillations tends to infinity.

\section{\label{sec:icosa} The resonators' locations}

Johnson and Merkowitz suggested that six resonator should be
positioned in the  center of the pentagonal faces of a truncated
icosahedron (TI) concentric to the sphere \cite{johnson1993}. The
TI configuration has several special symmetries. The pattern
matrix can be constructed in such a way that its pattern vectors
will be orthogonal in pairs. It has the following properties
\cite{merkowitz1997}
\begin{eqnarray}
  \underline{\underline{B}}\underline{\underline{B}}^T &=& \frac{3}{2\pi}\underline{\underline{I}} \\
  \underline{\underline{B}}\underline{1} &=& \underline{0}
\end{eqnarray}
and the following one, discovered during development of this work:
\begin{equation}
\underline{\underline{B}}^T\underline{\underline{B}}=\frac{3}{2\pi}\underline{\underline{I}}-\frac{1}{4\pi}\underline{\underline{1}}
\end{equation}
where $\underline{1}$ and $\underline{\underline{1}}$ are
respectively a vector and a matrix in which all elements are the
unity, and $\underline{0}$ represents the null vector.
\par These properties make the TI configuration a good option to the
resonator distribution on the sphere because it minimizes their
effects on each other while using a small number of resonators,
and make possible to obtain an analytic solution to the equations
of motion, as shown in section \ref{sec:moveqsol}.
\par Also, in TI configuration several resonators
are affected by a specific sphere mode. So the sphere modes can be
monitored by a linear combination of the six resonators outputs.
Jonhson and Merkowitz have named these linear combinations as mode
channels, $g_m (t)$, and we have adapted their definition as
\begin{equation}\label{eq:channelmodes}
\underline{g}(t)=\underline{\underline{B}}\underline{q_2}(t).
\end{equation}
They are a direct amplified reading of sphere modes.

\section{\label{sec:moveqsol} A solution for the equation of motion
and its analytic expression}

Equation \ref{eq:moveq} can be rewritten as
\begin{equation}\label{eq:moveq2}
\underline{\underline{X}}\underline{\underline{\gamma}}\underline{\ddot
w}(t)+ \omega_0^2
\underline{\underline{Y}}\underline{\underline{\gamma}}\underline{w}(t)
= \underline{\underline{Z}} \underline{F}(t),
\end{equation}
where
\begin{equation}
\underline{\underline{\gamma}}\underline{w}(t) \equiv \left(%
\begin{array}{ccc}
  \frac{1}{\sqrt{m_S}}\underline{\underline{I}} & \underline{\underline{0}} & \underline{\underline{0}} \\
  \underline{\underline{0}} & \frac{1}{\sqrt{m_{R_1}}} & \underline{\underline{0}} \\
  \underline{\underline{0}}& \underline{\underline{0}} & \frac{1}{\sqrt{m_{R_2}}} \\
\end{array}%
\right)\underline{w}(t)\equiv \left(%
\begin{array}{c}
  \underline{A}(t) \\
  \underline{q_1}(t) \\
  \underline{q_2}(t) \\
\end{array}%
\right) ,
\end{equation}
and $\underline{\underline{X}}$, $\underline{\underline{Y}}$ and
$\underline{\underline{Z}}$ are the original matrices from
equation \ref{eq:moveq} following their original positions. In
this way we solved the problem in a coordinate system based on
center of mass, such $\underline{\underline{\gamma}}$ is the
transformation matrix.
\par We can normalize the equation \ref{eq:moveq2} by multiplying it
to
$(\underline{\underline{X}}\underline{\underline{\gamma}})^{-1}$
through left and it became
\begin{equation}\label{eq:moveq3}
\underline{\ddot
w}(t)+\omega_0^2\underline{\underline{M}}\underline{w}(t) =
\underline{\underline{K}}\underline{F}(t),
\end{equation}
in which $\underline{\underline{M}}\equiv
(\underline{\underline{X}}\underline{\underline{\gamma}})^{-1}\underline{\underline{Y}}$
and $\underline{\underline{K}}\equiv
(\underline{\underline{X}}\underline{\underline{\gamma}})^{-1}\underline{\underline{Z}}$.

\par The next step is to obtain an orthogonal form of the equation
of motion to separate the $m+2j$ harmonic oscillators from each
other. That results in an equal number of linear equations and
simplify the  solution of the problem. We assume that
$\underline{\underline{M}}$ is symmetric (and it really is) and
therefore it can be diagonalized by using the relation
\begin{equation}\label{eq:diago}
\underline{\underline{U}}^\dag
\underline{\underline{M}}\underline{\underline{U}}=\underline{\underline{D}},
\end{equation}
where $\underline{\underline{U}}$ and $\underline{\underline{D}}$
represent respectively the matrix which columns are eigenvectors
and the eigenvalues diagonal matrix of
$\underline{\underline{M}}$. $\underline{\underline{D}}$ must have
positive and real values since frequencies must be real and
positive. Thus we impose that $\underline{\underline{U}}$ is
Hermitian
($\underline{\underline{U}}^\dag\underline{\underline{U}}=\underline{\underline{I}}$)
which assures $\underline{\underline{M}}$ is positively defined
and $\underline{\underline{D}}$ is positive and real. Therefore we
can rewrite \ref{eq:moveq3} as
\begin{equation}\label{eq:moveq4}
\underline{\ddot
\zeta}(t)+\omega_0^2\underline{\underline{D}}\underline\zeta(t)\underline{\underline{U}}^\dag\underline{\underline{K}}\underline{F}(t),
\end{equation}
where $\underline{\zeta}=\underline{\underline{U}}^\dag
\underline{w}$.
\par Now we can solve the equation of motion in the frequency domain where
equation \ref{eq:moveq4} becomes
\begin{equation}\label{eq:moveq5}
\left(-\omega^2\underline{\underline{I}}+\omega_0^2\underline{\underline{D}}
\right)\underline{\tilde{\zeta}}(\omega)=\underline{\underline{U}}^\dag\underline{\underline{K}}\underline{\tilde
F}(\omega).
\end{equation}
Since $\underline{\underline{D}}$ and $\underline{\underline{I}}$
are diagonal matrices the matrix sum in the parentheses is
invertible, implying that there is a matrix
$\underline{\underline{\tilde J}}^{-1}(\omega)$ that is equal to
that sum and we have
\begin{equation}\label{eq:solmoveq0}
\underline{\tilde\zeta}(\omega)=\underline{\underline{\tilde
J}}(\omega)\underline{\underline{U}}^\dag\underline{\underline{K}}\underline{\tilde
F}(\omega).
\end{equation}
To return to the original coordinates we reverse the
transformations:
\begin{equation}\label{eq:solmoveq1}
\left(%
\begin{array}{c}
  \underline{\tilde A} (\omega) \\
  \underline{\tilde q_1} (\omega) \\
  \underline{\tilde q_2} (\omega) \\
\end{array}%
\right)=\underline{\gamma}\underline{\underline{U}}
\underline{\tilde\zeta}(\omega)=\underline{\gamma}
\underline{\underline{U}}\underline{\underline{\tilde
J}}(\omega)\underline{\underline{U}}^\dag\underline{\underline{K}}\underline{\tilde
F}(\omega),
\end{equation}
which represents the solution of the equation of motion in the
frequency domain.

\subsection{\label{ssec:eigen} Eigenvalues and eigenvectors}

Looking at equation \ref{eq:solmoveq1} one can check that a good
determination of eigenvalues and eigenvectors of
$\underline{\underline{M}}$ is essential to solve the problem.
Moreover, the expressions must rigourously respect imposed
conditions. From numerical results for Schenberg's case we noticed
that $\underline{\underline{U}}$ can be separated into two
distinct groups:
\begin{eqnarray}\label{eq:eigenvectors}
\underline{U}_{1\pm}=n_{1\pm}\left(%
\begin{array}{c}
  \underline{0} \\
  \underline{1} \\
  d_{1\pm}\underline{1} \\
\end{array}%
\right)~~~\mathrm{and}\\
\underline{\underline{U}}_{k}=n_{k}\left(%
\begin{array}{c}
  \underline{\underline{I}} \\
  c_k\underline{\underline{B}}^T \\
  d_k\underline{\underline{B}}^T \\
\end{array}%
\right), ~\mathrm{for}~k=2,3,4,
\end{eqnarray}
where $\underline{\underline{U}}_{k}$ corresponds to three
degenerate frequencies and $\underline{U}_{1\pm}$ to isolated
frequencies, as is shown in subsection \ref{ssec:freq}. The
notation $\pm$ denotes eigenvectors which refer to up (+) and down
(-) shifting of the frequency from the central degeneracy. The
values of the constants $n_{1\pm}$, $d_{1\pm}$, $n_{k}$, $c_{k}$
and $d_{k}$ are obtained from Equation \ref{eq:diago} where
\begin{equation}
\underline{\underline{D}}=\left(%
\begin{array}{cc}
  \lambda_{1\pm} & \underline{0} \\
  \underline{0} & \lambda_k \underline{\underline{I}} \\
\end{array}%
\right),
\end{equation}
and they are shown in appendix \ref{ap:parameters}. Thus we have
\begin{equation}
\underline{\underline{\tilde J}}(\omega)=\left(%
\begin{array}{cc}
  \frac{1}{\omega^2-\omega_{1\pm}} & \underline{0} \\
  \underline{0} & \frac{1}{\omega^2-\omega_{k}}\underline{\underline{I}} \\
\end{array}%
\right),
\end{equation}
where
\begin{equation}
\omega_{1\pm}^2=\lambda_{1\pm}\omega_0^2 ~~\mathrm{and}~~
\omega_{k}^2=\lambda_{k}\omega_0^2,
\end{equation}
correspond to the resonant frequencies of the system.

\subsection{\label{ssec:freq} The coupled mode frequencies}

\begin{figure*}\label{fig:freq}
  \includegraphics[width=13cm]{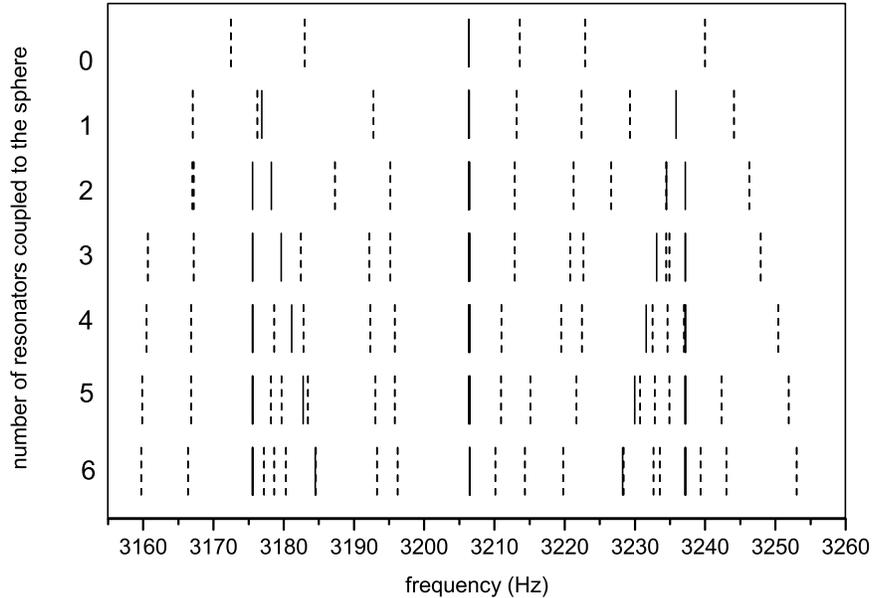}\\
  \caption{Frequencies of the coupled modes.}
\end{figure*}

\begin{figure}\label{fig:distr}
  \includegraphics[width=6cm]{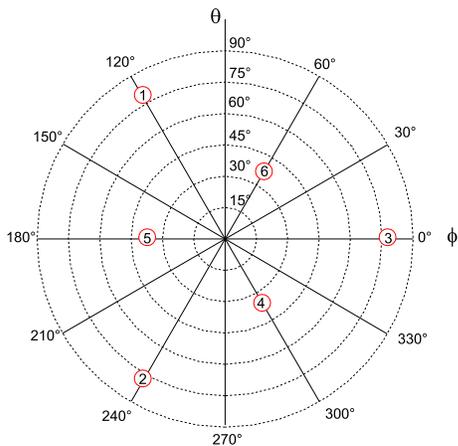}\\
  \caption{Projected position of the resonators on the sphere's surface.}
\end{figure}

Our tests for many $i$-mode identical resonators coupled to the
sphere using TI configuration results that eigenfrequencies
present $i+1$ degenerated quintuplets and $i$ non-degenerated
modes. In Mario Schenberg's case we have three degenerate
quintuplets ${(f \sim 3175.6, 3206.4, 3237.2\mathrm{Hz})}$ and two
isolated modes ${(f \sim 3184.5, 3228.3 \mathrm{Hz})}$. Figure
\ref{fig:freq} shows how resonance frequencies change as more
transducers are coupled to the antenna. The number on the left
corresponds to the number of resonators gradually added following
the ordering in Figure \ref{fig:distr}: the number zero
corresponds to the isolated sphere; the number one corresponds to
resonator 1 coupled to the sphere; the number two corresponds to
resonators 1 and 2 coupled to the sphere and so forth. Dashed
lines in the results with zero resonators indicate the frequencies
measured of the real sphere at $4.2\mathrm{K}$.
\par In Figure \ref{fig:freq} we present how the resonant
frequencies should behave when we consider non-degenerate sphere
modes, like those resultant from the measurements performed (the
dashed lines we mentioned above). We concluded that there is a
negligible effect from the suspension hole on the symmetry
proprieties of system so we decided to consider degenerate initial
modes as a good approximation.

\subsection{\label{ssec:analytic} An analytic expression to the solution of
the  equation of motion}

The special symmetries presented by the TI configuration enable us
to find an analytic expression to the solution of the equation of
motion \cite{magalhaes1997}. After a lengthy calculation we found
an analytic expression for the mode channels, given by
\begin{eqnarray}\label{eq:modechannels}
\underline{\tilde
g}(\omega)&=&\frac{3}{2\pi}\frac{1}{\sqrt{{m_R}_2}}\sum_{k=2}^4
\eta_k d_k \lambda_k \underline{\tilde F}_S
(\omega)\nonumber\\
&+&\frac{3}{2\pi}\frac{\alpha}{b\sqrt{m_{R_2}}}
q\sum_{k=2}^4\eta_k d_k \left(c_k
-\frac{b+a}{b}\right)\underline{\underline{B}}\underline{\tilde
{F_1}}^N\nonumber\\
&+&\frac{3}{2\pi}\frac{\alpha}{b\sqrt{m_{R_2}}}\sum_{k=2}^4\eta_kd_k
\left[c_k+d_k(a+a^{-1})\right]\underline{\underline{B}}\underline{\tilde
{F_2}}^N,\nonumber\\
\end{eqnarray}
where $$q=\left(\frac{1}{b}+\frac{3}{2\pi}b\right),$$
$$a=\sqrt{\frac{m_{R_2}}{m_{R_1}}},$$ and $$b=\alpha\sqrt{\frac{m_{R_2}}{m_S}}.$$
The value of $\eta_k$ depends of $n_{k}$, $c_{k}$ and $d_{k}$
values and it also is shown in appendix \ref{ap:parameters}.
\par The above equation results from a combination of equations
\ref{eq:channelmodes} and \ref{eq:solmoveq1}. It represents the
solution of the equation of motion and allows us to know the
system's behavior under a gravitational wave excitation.

\section{\label{sec:simul} Numerical simulation of the system}

\begin{figure}
\includegraphics[width=8.5cm]{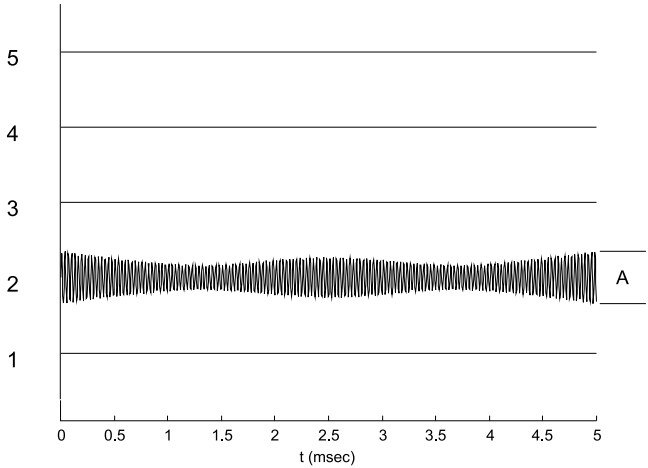}\\
  \caption{\label{fig:At} Responses of the sphere's normal modes to
  the simulated gravitational wave.
  Only the second normal mode was excited, as expected.}
\end{figure}

In order to test the model we simulated a signal as a
gravitational sinewave quadrupolar force with frequency equal to
the degenerate normal mode frequencies of the sphere ${(\sim
3206.3 \mathrm{Hz})}$. The signal would be arriving at the local
zenith so that its propagation direction coincided with the $z$
axis of the reference system centered at SCM. Also, we assumed
that the simulated gravitational wave had only ``$\times$"
polarization (``$+$" polarization was null). Those aspects made
only the second normal mode to be excited by the passing of the
wave, as shown in Figure \ref{fig:At}. We ignored any noise
components in this simulation ($F_i^N=0$).
\begin{figure}
\includegraphics[width=8.5cm]{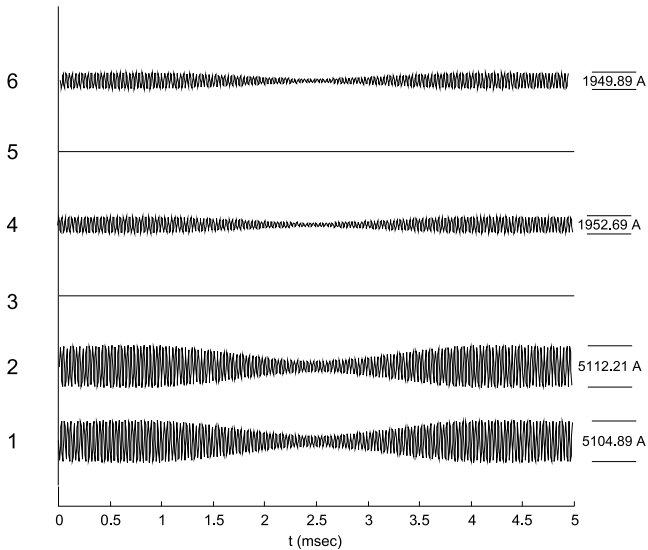}\\
  \caption{Amplitudes of the second resonators ($R_2$).}\label{fig:q2}
\end{figure}
\par Special symmetries presented by TI configuration allow only
some resonators to oscillate when only the second normal mode is
excited, as shown in Figure \ref{fig:q2}. The number on the right
represent relative amplification to maximum amplitude of normal
modes of the sphere.
\par
\begin{figure}
\includegraphics[width=8.5cm]{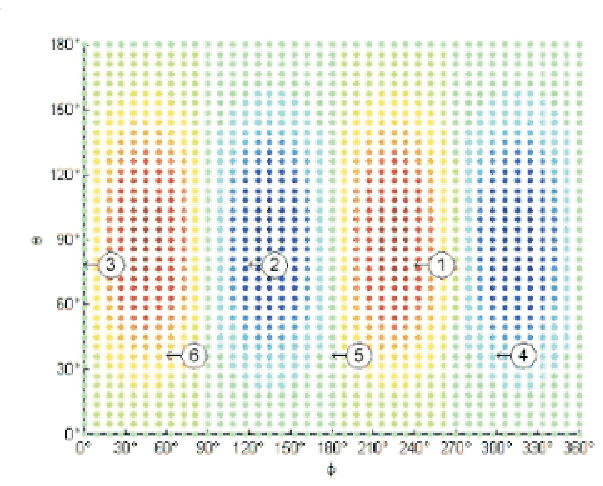}\\
  \caption{Location of the transducers compared with second
  normal mode of the sphere. The darkest areas represent maximum
  radial amplitude loci.}\label{fig:planmodes}
\end{figure}

 Richard showed (using the Principle of Energy Conservation)
that the ratio of displacements \cite{richard1982}
\begin{equation}\label{eq:radisp}
\frac{|x_2|}{|x_1|}\equiv \sqrt{\frac{m_1}{m_2}}.
\end{equation}
We apply this concept to Schenberg's case and found
\begin{equation}\label{eq:radisp2}
\frac{|q_2|}{|A|}\equiv\sqrt{\frac{m_{eff}}{m_{R_2}}} \simeq 5367,
\end{equation}
in resonant frequencies. This value is higher than that shown in
Figure \ref{fig:q2}. However as one can observe in Figure
\ref{fig:planmodes} no resonator is on a region of maximum
amplitude which justifies the difference between the values.
Figure \ref{fig:planmodes} presents a plane representation of
second normal mode of sphere where darkest areas correspond to
maximum displacement loci. One can observe the resonator 3 and 5
are placed on null radial movement areas. Resonators 1 and 2 are
placed close of maximum displacement points. An analogous analysis
are done for all others normal modes and compatible results are
obtained.

\begin{figure}
\includegraphics[width=8.5cm]{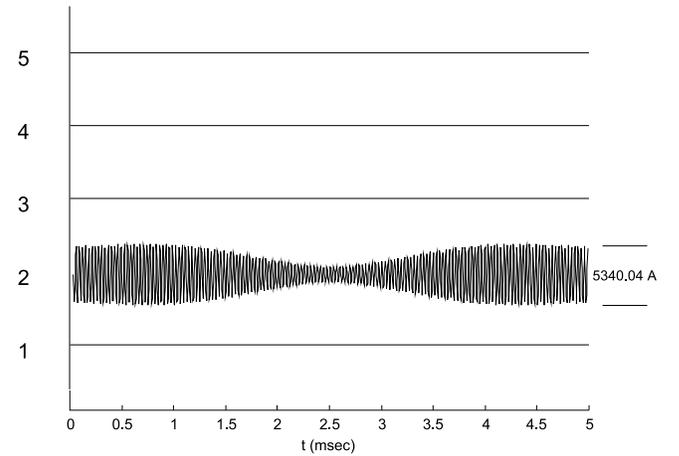}\\
  \caption{Mode channel outputs for the simulated signal.}\label{fig:g}
\end{figure}

From equation \ref{eq:channelmodes} we obtain the mode channels
outputs as showed in Figure \ref{fig:g}.

\section{An estimate of the noise contribution on the mode channels}

We can rewrite Equation \ref{eq:modechannels} as
\begin{equation}\label{eq:modechannels2}
\tilde{g}_m(\omega)=\xi(\omega)\tilde{F}^S_m+\Omega_{mj}^{R_1}(\omega)\tilde{F}^N_{j1}+\Omega_{mj}^{R_2}(\omega)\tilde{F}^N_{j2}
\end{equation}
where $\xi(\omega)$ corresponds to the transfer function of the
$\tilde F^S_m$ components on $g_m$. Similarly, the
$\Omega_{mj}^{R_i}(\omega)$ matrices represent the response
functions of   mode channel $m$ to the noise source $\tilde
F^N_{ji}$.
\par Assuming  statistically independent noise sources and
ignoring $\mathcal{\tilde F}_m$ on $\tilde F^S_m$, the mode
channels spectral densities , $S^g_m (\omega)$, can be written as
\begin{eqnarray}\label{eq:specden}
S^g_m (\omega)=|\xi(\omega)|^2S^{\tilde{F}^S_m}+
\sum_j|\Omega_{mj}^{R_1}(\omega)|^2S^{\tilde{F}^N_{j1}}+
\nonumber\\\sum_j|\Omega_{mj}^{R_2}(\omega)|^2S^{\tilde{F}^N_{j2}},
\end{eqnarray}
with $S^{\tilde{F}^S_m} \equiv \tilde{F}^S_m \ast \tilde{F}^S_m$
and $S^{\tilde{F}^N_{ji}} \equiv \tilde{F}^N_{ji} \ast
\tilde{F}^N_{ji}.$ As an example one can calculate the noise
contribution from Langevin forces, which have spectral densities
well known and given by
\begin{equation}\label{eq:Langevin}
S^{FL_i}=\frac{4 k_B T m_i \omega}{Q_i},
\end{equation}
where $k_B=1.38\times 10^{-23}\mathrm{J/K}$ is Boltzman's
constant, $T$ is physical temperature of the system, and $m_i$ and
$Q_i$ are respectively the physical mass and the mechanical
quality factor of the body $i$.
\par
\begin{figure}\label{fig:tese6_13}
\includegraphics[width=8.5cm]{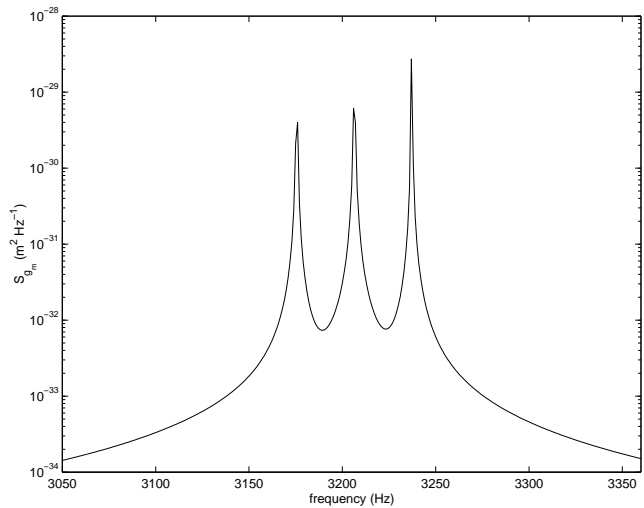}\\
  \caption{Spectral density of the contributions of the Langevin forces
  to the mode channels.}
\end{figure}

 From the calculation of the of the spectral density $
S_m^g \left( \omega  \right)$, assuming no incident GW signal, we
found that the contributions to each normal mode seem to have the
same spectral densities, but with  slightly different values,
since each one of them receives different contributions from each
of the resonators (See Figure \ref{fig:tese6_13}).
These contributions are small compared to those originated from
the sphere since the resonators' masses are much smaller than the
sphere's mass ($
 {{m_s } \mathord{\left/
 {\vphantom {{m_s } {Q_s }}} \right.
 \kern-\nulldelimiterspace} {Q_s }}\sim 34{{m_{R1} } \mathord{\left/
 {\vphantom {{m_{R1} } {Q_{R1} }}} \right.
 \kern-\nulldelimiterspace} {Q_{R1} }}\sim 5.7 \times 10^3 {{m_{R2} } \mathord{\left/
 {\vphantom {{m_{R2} } {Q_{R2} }}} \right.
 \kern-\nulldelimiterspace} {Q_{R2} }}$,
using the adopted values). The agreement among the values of
$S_m^g \left( \omega  \right)$ is also a consequence of the fact
that we have initially assumed that all the modes have the same
transfer function, $ \xi \left( \omega  \right) $. The use of
different transfer functions for each mode $m$ may help the
understanding on how asymmetries on the real sphere should affect
the individual responses of the modes to GWs \cite{stevenson1996}.
\par
\begin{figure}\label{fig:sensitivity}
\includegraphics[width=8.5cm]{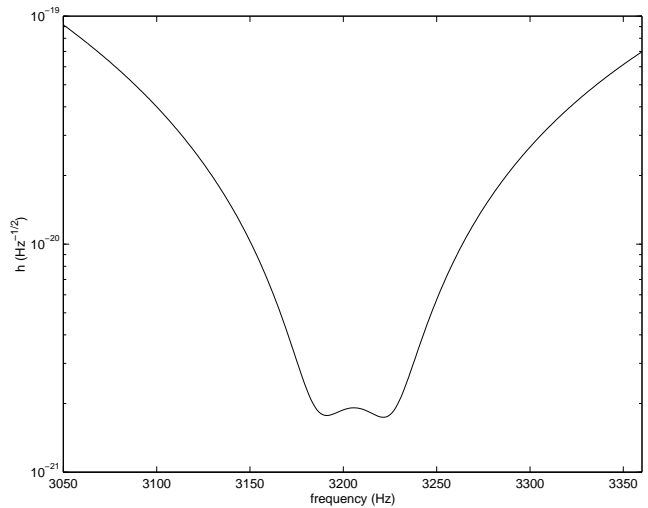}\\
  \caption{Sensitivity curve for the mode channels.}
\end{figure}

 The GW effective force (eq. \ref{eq:egf}) that acts on the sphere
(in the frequency domain) has spectral density
\begin{equation}\label{eq:SFm}
S^{\mathcal{F}_{GW}}_m (\omega)=\left(\frac{1}{2}\omega^2m_S\chi R
\right)^2 S^{\tilde h}_m(\omega).
\end{equation}
The sensitivity curve for mode $m$ is given through
\begin{equation}\label{eq:hm}
\tilde{h}_m(\omega)=\sqrt{\frac{1}{\left(\frac{1}{2}\omega^2m_S\chi
R \right)^2}\frac{S^{g}_m(\omega)}{|\xi_m(\omega)|^2}}.
\end{equation}
Figure \ref{fig:sensitivity} shows the sensitivity curves for the
five quadrupolar normal modes of the sphere. We obtained this
curve assuming that no signal was present and that only the usual
noises in this kind of detector (Brownian noise, serial noise,
back-action noise and electronic phase and amplitude noises) were
disturbing the antenna.
\par From reasonable estimates of the spectral densities of the
noise sources it is possible to determine their contributions to
the mode channels and, consequently, the values of $ \tilde h_m
\left( \omega  \right)$, which allow for the location of the GW
source in the sky, as well as the determination of the
polarization amplitudes of the GW.

\section{Conclusion}

In this work we presented a model that describes the response to a
gravitational wave of the spherical GW detector SCHENBERG coupled
to six one-dimensional two-mode transducers .
\par The analytical expressions
obtained as solutions of the equation of motion for the system are
in good agreement with the numerical solutions.
\par We found an expression (Equation \ref{eq:specden}) which
allows for estimates of the contributions of the noises to the
mode channels. Such an expression is very much needed for a more
realistic modelling of the SCHENBERG detector.

\appendix
\section{Eingenvector solutions}\label{ap:parameters}
In order to determine the values of the constants $n_k, c_k, d_k,
n_{1\pm}$ and $d_{1\pm}$ we imposed that
$\underline{\underline{U}}$ must be Hermitian and
\begin{equation}\label{eq:diageq}
\underline{\underline{M}}\underline{\underline{U}}=\underline{\underline{U}}\underline{\underline{D}}.
\end{equation}
We separate the two groups of eigenvectors ($\underline{U}_{1\pm}$
and $\underline{\underline{U}}_k$) and solve equation
\ref{eq:diageq} using
\begin{equation}\label{eq:eingsol1}
\underline{\underline{M}}\underline{U}_{1\pm}=\underline{U}_{1\pm}D_{1\pm},
\end{equation}
\begin{equation}\label{eq:eingsol2}
\underline{\underline{M}}\underline{\underline{U}}_{k}=\underline{\underline{U}}_{k}D_{k},
\end{equation}
where $D_{1\pm}\equiv\lambda_{1\pm}$ and $D_{k}\equiv\lambda_k$
are the eingenvalues of eigenvectors $\underline{U}_{1\pm}$ and
$\underline{\underline{U}}_k$ respectively. So, by using the
properties  of the $\underline{\underline{B}}$ matrix we obtain
the following expressions for the constants:
\begin{equation}\label{eq:n1}
 n_{1\pm}^2 = \frac{1}{1+\frac{3}{2\pi}\left(1+d_{1\pm}^2\right)},
\end{equation}
\begin{equation}\label{eq:d1}
 d_{1\pm} = -\frac{1}{2}\left(a\pm\sqrt{a^2+4}\right),
\end{equation}
\begin{equation}\label{eq:nk}
n_k^2=\frac{1}{1+\frac{3}{2\pi}(c_k^2+d_k^2)},
\end{equation}
\begin{equation}\label{eq:dk}
d_k=\frac{3}{2\pi}\frac{b}{a}\left(c_k^2+bc_k-\frac{2\pi}{3}\right),
\end{equation}
\begin{equation}\label{eq:c2}
c_2=\frac{s_3^2+4x^2-2xs_3-2y}{6s_3},
\end{equation}
\begin{eqnarray}\label{eq:c34}
  c_{3,4} &=& \frac{4(-1\mp i\sqrt{3})x^2+(-1\pm i\sqrt{3})s_3^2-4xs_3}{12s_3}\nonumber\\
  &+& \frac{1\pm
i\sqrt{3}}{s_3},
\end{eqnarray}
where
\begin{equation}\label{eq:imag}
i=\sqrt{-1},
\end{equation}
\begin{equation}\label{eq:s3}
s_3=(s_1+12\sqrt{s_2})^{1/3},
\end{equation}
\begin{equation}\label{eq:s1}
s_1=-8sx^3+36xy-108z,
\end{equation}
\begin{equation}\label{eq:s2}
12y^3+12x^3z-3x^2y^2+81z^2-54xyz,
\end{equation}
\begin{equation}\label{eq:x}
x=\frac{2\pi}{3}\left(\frac{a^2}{b}+\frac{3}{2\pi}b\right),
\end{equation}
\begin{equation}\label{eq:y}
y=\frac{2\pi}{3}\left[\frac{a^2}{b^2}\left(b^2-\frac{2\pi}{3}\right)-1\right]
\end{equation}
\begin{equation}\label{eq:z}
z=-\frac{4\pi^2}{9}\frac{a^2}{b}.
\end{equation}
Those imply that
\begin{equation}\label{eq:lambda1}
\lambda_{1\pm}=1-ad_{1\pm}
\end{equation}
and
\begin{equation}\label{eq:lambdak}
\lambda_{k}=1-\frac{3}{2\pi}b^2c_k.
\end{equation}

Although $c_{3,4}~(c_k~\mathrm{with}~k=3,4)$ has imaginary terms
its value is real, which satisfies Hermitian condition for
eigenvectors.
\par We also can define the quantities
\begin{equation}\label{eq:eta1}
\eta_{1\pm}=\eta_{1\pm}(\omega)=\frac{n_{1\pm}}{\omega_{1\pm}^2-\omega^2}
\end{equation}
and
\begin{equation}\label{eq:etak}
\eta_{k}=\eta_{k}(\omega)=\frac{n_{k}}{\omega_{k}^2-\omega^2}.
\end{equation}
These quantities help us to write an analytic expression for the
mode channel outputs.

\begin{acknowledgments}
This work was supported by \textbf{FAPESP} (under grant numbers
1998/13468-9, 2001/14527-3 and 2003/02912-5), \textbf{CNPq} (under
grant number 300619/92-8) and \textbf{MCT/INPE}.
\end{acknowledgments}

\bibliography{Schenberg}

\end{document}